\documentclass[fleqn,usenatbib]{mnras}

\usepackage{newtxtext,newtxmath} 
\usepackage[T1]{fontenc}
\usepackage{ae,aecompl} 
\usepackage{graphicx}	
\usepackage{amsmath}	
\usepackage{amssymb}	

\newcommand{\om}{\Omega_{\rm m}} 
\newcommand{\lcdm}{$\Lambda$CDM} 

\newcommand{\be}{\begin{equation}}
\newcommand{\ee}{\end{equation}}
\newcommand{\bs}{\begin{split}} 
\newcommand{\bea}{\begin{eqnarray}}
\newcommand{\eea}{\end{eqnarray}}

\title[Gravitational Waves Determine Hubble Constant?]{Will Gravitational Wave Sirens Determine the Hubble Constant?} 

\author[Shafieloo et al.]{
Arman Shafieloo,$^{1,2}$\thanks{Email: shafieloo@kasi.re.kr}
Ryan E.\ Keeley,$^{1}$\thanks{E-mail: rkeeley@kasi.re.kr}
Eric V.\ Linder$^{1,3,4}$\thanks{E-mail: evlinder@lbl.gov} 
\\
$^{1}$Korea Astronomy and Space Science Institute, Daejeon 34055, Korea\\
$^{2}$University of Science and Technology, Yuseong-gu 217 Gajeong-ro, Daejeon 34113, Korea\\
$^{3}$Berkeley Center for Cosmological Physics \& Berkeley Lab,
University of California, Berkeley, CA 94720, USA\\
$^{4}$Energetic Cosmos Laboratory, Nazarbayev University, Astana, Kazakhstan 010000
}

\date{Accepted XXX. Received YYY; in original form ZZZ}

\pubyear{2018}

\begin{document}
\label{firstpage}
\pagerange{\pageref{firstpage}--\pageref{lastpage}}
\maketitle

\begin{abstract} 
Lack of knowledge about the background expansion history of the Universe from independent observations makes it problematic to obtain a precise and accurate estimation of the Hubble constant $H_0$ from gravitational wave standard sirens, even with electromagnetic counterpart redshifts. 
Simply fitting simultaneously for the matter density in a flat \lcdm\ model 
can reduce the precision on $H_0$ from 1\% to 5\%, while not knowing the actual background expansion model of the universe (e.g.\ form of dark energy) can introduce substantial bias in estimation of the Hubble constant. 
When the statistical precision is at the level of 1\% uncertainty on $H_0$, biases in non-\lcdm\ cosmologies that are consistent with current data could reach the 3$\sigma$ level. To avoid model-dependent biases, statistical techniques that are appropriately agnostic about model assumptions need to be employed.

\end{abstract}




\section{Introduction}

Gravitational waves emitted from inspiral and coalescence of binary compact objects can be used to measure a dimensional quantity -- the time or frequency associated with the wave form. By modeling the expected wave form within general relativity these events can be standard sirens, measuring  dimensional cosmic distances. Since most cosmic measurements involve dimensionless quantities (often ratios of distances), this makes standard siren distances potentially useful in a distinct way. In particular, they have been proposed to measure the absolute distance scale of the universe, or Hubble constant $H_0$ \citep{Schultz:1986,2005ApJ...629...15H,2006PhRvD..74f3006D}. This is an exciting prospect. 

Locally, at very low redshifts $z\lesssim0.1$, the source distance is related linearly to 
the distance through the Hubble law, $d=H_0^{-1}z$. This means that the redshift to the 
source must also be determined, but it is not uniquely provided by the gravitational wave 
(GW) observations. The most straightforward way to obtain the redshift is to use GW systems 
with electromagnetic (EM) counterpart events (e.g.\ X-ray or optical flashes associated 
with the merger), where the redshift comes from the EM measurement. (Crosscorrelation with redshift surveys is an 
alternative area of investigation, e.g.\ \cite{Zhang};
see \cite{Aasi:2013wya} for a general review of GW detectors). 

However, low redshift means a small volume in which events can occur, hence small numbers 
of observed GW+EM systems and poor precision on $H_0$. Upcoming gravitational wave 
detectors -- more sensitive runs of LIGO-Hanford and LIGO-Livingston \citep{TheLIGOScientific:2014jea}, 
Virgo \citep{TheVirgo:2014hva}, and new interferometers in India and Japan \citep{india,2013PhRvD..88d3007A} -- 
will be able to detect GW events expected to have EM counterparts (e.g.\ binary neutron 
stars) out to higher redshift, $z\approx0.5$, with further 
generations including space based detectors such as LISA \citep{2013GWN.....6....4A,2017arXiv170200786A} reaching $z\approx1$ 
or beyond. These will detect significantly more events, and some papers have 
held them out as the means to measure the Hubble constant to 1\% or better precision (see, e.g., \citet{Chen}).

Obtaining a 1\% measurement of $H_0$ from low redshift observables can be important for cosmology because of the discordance between the inference of $H_0$ from observations of the cosmic microwave background (CMB) \citep{2018arXiv180706209P} and the value observed locally from calibrating the supernova (SN) distance ladder with Cepheids \citep{2016ApJ...826...56R}.  An independent measurement of $H_0$ would offer keen insight into this tension (e.g.\ \cite{2018arXiv180203404F}).  If $H_0$ from GW preferred the Planck value, then that might indicate an unaccounted for systematic with Cepheid calibrations of SN distances.  If it agrees with the Cepheid+SN measurement, then that might lend more credence to the idea this discordance is explained by new physics.  

At redshifts $z\gtrsim0.1$, though, the distance does not depend solely on $H_0$ but on an integral over the expansion history $H(z)$, with all components of energy density in the universe -- in particular matter and dark energy -- contributing. Without knowledge of these components one cannot cleanly separate out the Hubble constant. 

Astrophysical systematics such as binary orbit inclination effects, peculiar velocities, and signal to noise (Malmquist-like) biases can also affect the use of GW sirens to constrain $H_0$ (see, e.g., \citet{2018arXiv181111723M}). For local sirens this could include coherent velocity flows (cf.\ 
\citet{2006PhRvD..73l3526H,2006PhRvD..73j3002C}). 

Our focus here is two-fold:  investigation of the precision with which $H_0$ can be constrained in a background more general than \lcdm\ -- and indeed whose form may be unknown, and investigation of the accuracy, i.e.\ the bias suffered when the background is not accounted for correctly. 
\citet{Chen} have put forth the exciting prospect of 1-2\% measurement of $H_0$ from GW but within a calculation assuming not only \lcdm\ but a 
perfectly known matter density. 
While \cite{2018PhRvD..98h3523D} have looked beyond a \lcdm\ background, this was only for non-GW probes, implementing the GW data as purely a $H_0$ prior from \cite{Chen}.  \cite{2018arXiv181201440D} have included a full dynamical dark energy background, but for far future GW data sets of 1000 sirens out to $z=5$  and in combination with other probes already giving strong cosmology constraints. 
Our approach is to examine the role of the background expansion model 
on both the precision and accuracy of $H_0$ determination from mid and moderately long term GW experiments. 

In Sec.~\ref{sec:data} we present the framework for the analysis, including the GW+EM datasets corresponding to next, and next next, generation GW experiments and our simulation methodology. 
Precision on $H_0$ is treated in Sec.~\ref{sec:bloat}, where we examine how it degrades with greater freedom for the expansion history: first including just the matter density and a cosmological constant, then allowing for dynamical dark energy with assumption of the standard $w_0$--$w_a$ time dependence. 
Accuracy is the focus in Sec.~\ref{sec:bias}, where we show how assuming a $\Lambda$CDM cosmology can significantly bias the results in the regime of 1\% precision.  In Sec.~\ref{sec:concl} we conclude and discuss the statistical techniques needed to infer $H(z)$ without bias from high precision datasets.

\section{Model Assumptions and Data Simulations} \label{sec:data}  

As stated in the Introduction, GW standard sirens do not directly measure $H_0$ but rather measure luminosity distances $D_L$ throughout the cosmic volume to which the detectors are sensitive. Each event has a cosmic redshift associated with it, which must be obtained from EM counterparts. The distance is then related to the cosmic expansion rate through 
\begin{equation}\label{eq:DL}
    D_L(z) = (1+z)\frac{c}{H_0} \int_0^z \frac{dz'}{h(z')}\ ,
\end{equation}
where $h(z)$ is the Hubble rate scaled to the present value, $H(z)/H_0$, and a spatially flat universe is assumed. The assumptions about the background expansion model $h(z)$ have a direct impact on extraction of $H_0$. Even under the assumption of flat \lcdm\ the matter density must also be known: $h^2(z)=\om(1+z)^3+1-\om$. 

To be concrete about how uncertainties in the expansion history can affect model dependent inferences using GWs, we generate mock luminosity distance datasets from a given background cosmology. We consider two alternative possibilities: 1) a flat \lcdm\ model with $\om=0.3$ and $h\equiv h(z=0)=0.69$, and 2) time varying dark energy with assumption of $w(z)=w_0+w_a z/(1+z)$ which provides more flexibility to the expansion history.


For the latter we choose two models consistent with current data, specifically lying on the 68\% confidence contour of the Pantheon supernovae plus Planck CMB plus SDSS BAO plus HST $H_0$ combined data fit of \cite{2018ApJ...859..101S}. 

To generate a realistic mock dataset of GW luminosity distances, we sample the event redshift 
distribution based on the assumption that the GW events have a constant rate per comoving volume. That is, 
\begin{equation}
    N(z)=\frac{dN}{dz} = \frac{dN}{dV_c}\frac{dV_c}{dz},
\end{equation}
where we assume $dN/dV_c$ is  constant and calculate $dV_c/dz$ from our fiducial input cosmology.  

We perform this sampling for two cases. One is a next generation case corresponding roughly to the $\sim$2026 HLVJI array of Hanford and Livingston detectors of LIGO, and Virgo, KAGRA (Japan), and LIGO-India detectors. We use a maximum redshift $z=0.5$ and draw 120 events from this distribution, with distance errors normally distributed with a $1\sigma$ precision of 13\%, roughly following \cite{Chen}.  

\begin{figure}
    \centering
    \includegraphics[width=\columnwidth]{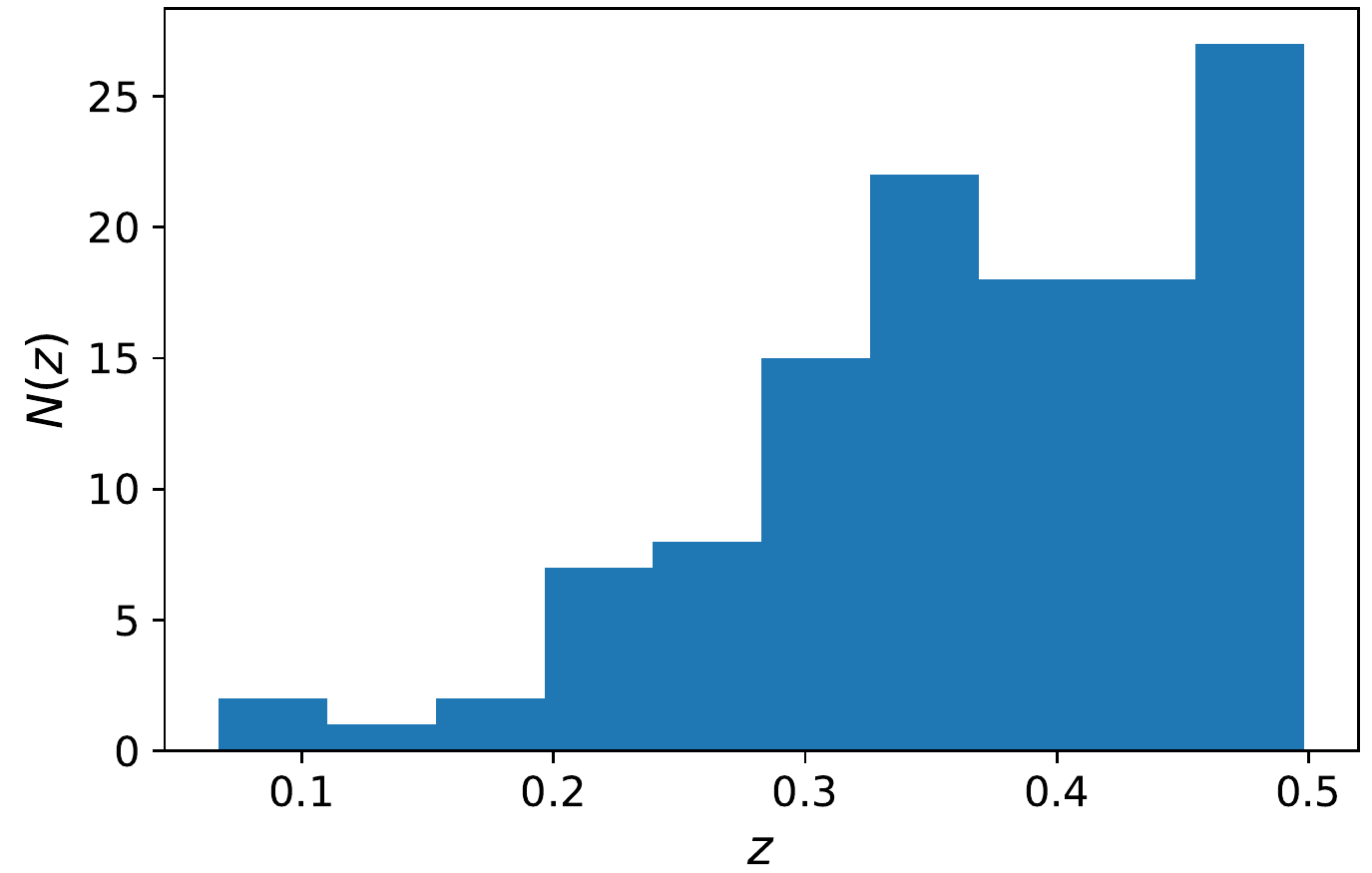}
    \includegraphics[width=\columnwidth]{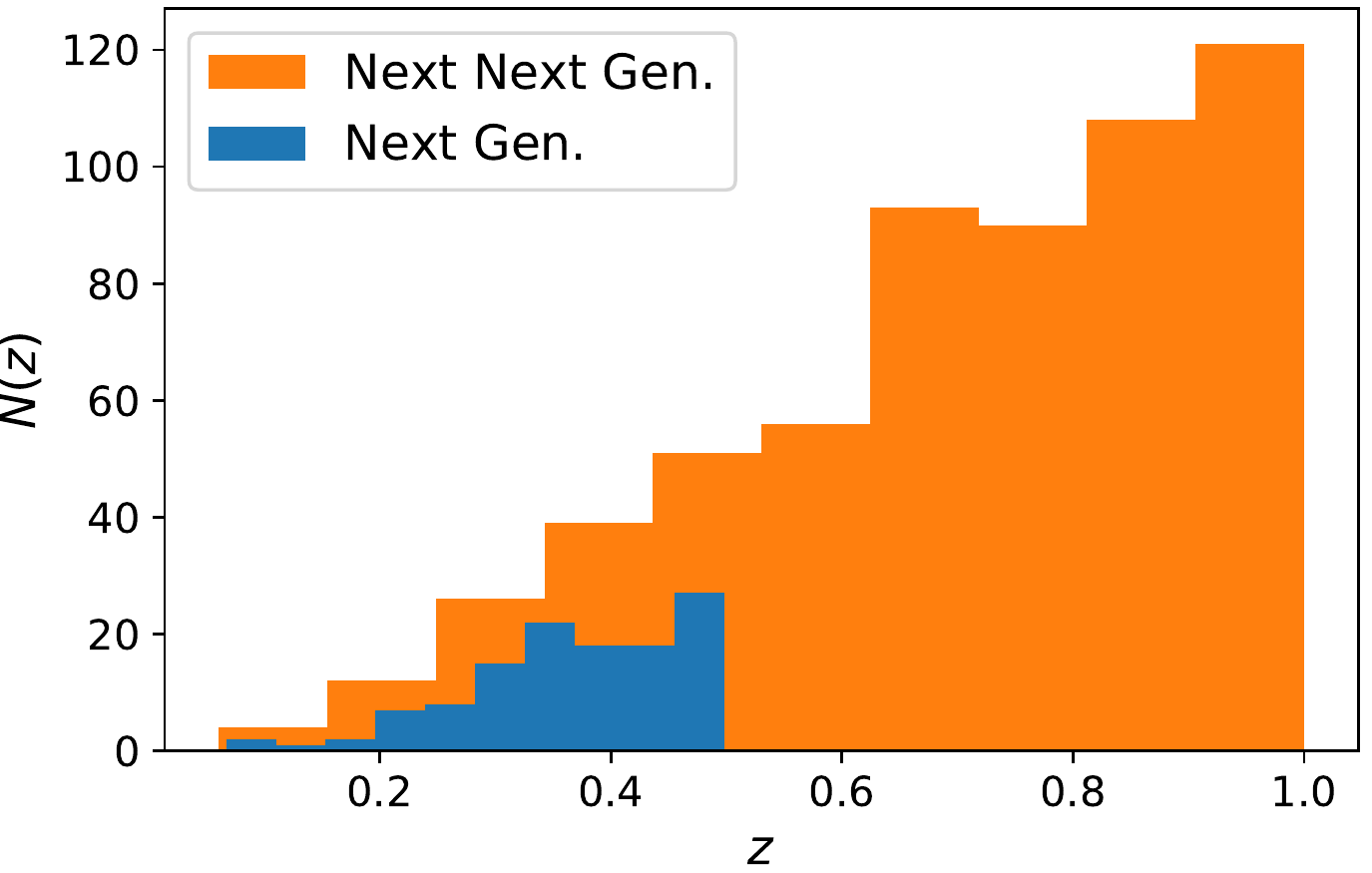}
    \caption{One realization of the  redshift distribution of GW+EM events for the next generation case of 120 total events observed out to maximum redshift  $z=0.5$ (upper panel) and for the next next generation case of 600 total events observed out to maximum redshift of $z=1.0$ (lower panel). The next generation events are overplotted on the lower panel for comparison.}
    \label{fig:zsample}
\end{figure}

The other is a next next generation case where we take the maximum redshift to be $z=1.0$ and draw 600 events from the distribution, with 7\% distance precision. One realization each of these samplings can be seen in Fig.~\ref{fig:zsample}.

Again we emphasize that any projection of the distance information from these redshifts to $H_0=H(z=0)$ requires the assumption of an uncertain background model.  Testing whether or not this model is true is of course one aim of any analysis of cosmological datasets, along with $H_0$. 


\section{Precision vs Cosmological Model} \label{sec:bloat} 

Combining multiple cosmological probes within a given model can give tight constraints on parameters, including $H_0$. As a rough rule of thumb, note that the CMB already tightly constrains 
the combination $\om h^3$, to about 0.3\% \citep{2018arXiv180706209P}, fairly independently of late time physics (though still power-law form of the primordial power spectrum is assumed). Thus $\delta h/h\approx (1/3)\delta\om/\om$ so a prior of 0.03 on $\om$ from large scale structure probes gives a 3\% constraint on $h$. Adding other probes such as supernovae would tighten this further. Thus we want another, individual probe at the level of $\sim 1\%$ on $h$. Let us explore under what 
conditions GW sirens can provide this in themselves.

\begin{table} 
\begin{center} 
\begin{tabular}{l|l|l|l} 
Background & Prior & $\sigma(h)$ \ & $\sigma(h)/h$ \\ 
\hline  
\lcdm\ & none & 0.036 & 5\% \\ 
\lcdm\ & $\sigma(\om)=0.03$\ & 0.010 & 1.4\% \\ 
\lcdm\ & fix $\om$\ & 0.0083\ & 1.2\% \\ 
$w_0$--$w_a{}^*$ & none & 0.039 & 6\% \\ 
$w_0$--$w_a{}^*$ & $\sigma(\om)=0.03$ & 0.039 & 6\% \\
$w_0$--$w_a{}^*$ & fix $\om$\ & 0.039 & 6\% \\ 
\end{tabular}
\end{center}
\caption{Constraints on $H_0$ from the next generation GW+EM set are given under various backgrounds and priors. An asterisk denotes a broad prior of 1 on both $w_0$ and $w_a$, since the Fisher information approach is inaccurate for extended degeneracies. We see that GW siren constraints are sensitive to the input model. 
} 
\label{tab:H0priors} 
\end{table}

Table~\ref{tab:H0priors} shows the constraints on $H_0$ from next generation GW data, either alone or with external priors. For this table alone, the numbers come from Fisher information computation; all other numbers in the article are from MCMC. They are in good agreement where 
they overlap. When the background is fixed to \lcdm, and furthermore the matter density is perfectly known, then the uncertainty is $\sigma(h)=0.0083$ or approximately 1.2\%. When external information is used at the level of a prior on matter density of $\sigma(\om)=0.03$ then the $H_0$ precision has a modest increase to 1.4\%. However, such combination of probes can be done as well between non-GW probes (e.g.\ supernovae, strong lenses, large scale structure, CMB), so 
we should look at what GW sirens themselves deliver. For GW alone, even with fixing the background to \lcdm, the uncertainty is 
$\sigma(h)=0.036$ or 5\%. This will not allow them to make a statistically significant statement on the tension between the Planck value (which used the same assumption of \lcdm) and the local distance ladder value. 

\begin{figure}
    \centering
    \includegraphics[width=\columnwidth]{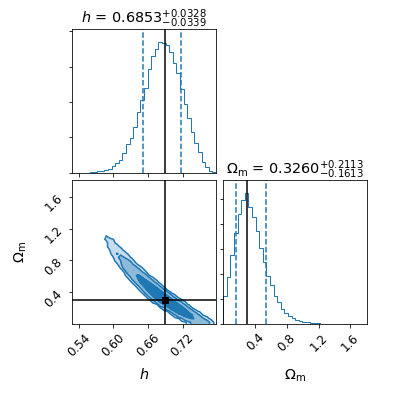}
    \includegraphics[width=\columnwidth]{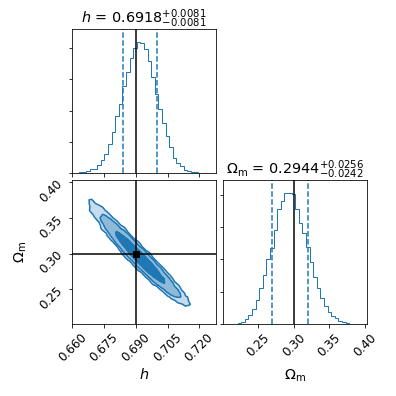}
    \caption{Forecast posterior for mock data generated from a $\Lambda$CDM cosmology for next generation (top) and next next generation (bottom) sensitivities.
    The 2D posterior for $h$ and $\Omega_{\rm m}$ shows the 68.3\%, 95.4\%, 99.7\% 
    confidence regions in increasingly lighter shades of blue.  The 1D posteriors show the 68.3\%  confidence regions in dashed blue.  The input values for $\Omega_{\rm m}$ and $h$ are indicated with solid black lines. 
    }
    \label{fig:LCDM_nearfuture}
\end{figure}

To illustrate the effect of the matter density covariance with $H_0$ within the \lcdm\ model, we perform a Monte Carlo simulation of next generation GW data. That is, we generate realizations 
of the data in a particular \lcdm\ case with $\om=0.3$ and $h=0.69$, 
and fit for these two parameters under the assumption we know the 
background is \lcdm. The 1D and 2D joint confidence contours 
appear in Fig.~\ref{fig:LCDM_nearfuture}.

The covariance between $\om$ and $h$ is clear, showing that -- 
just as with other distance probes -- external data to break 
parameter degeneracies is necessary. Only if $\om$ is well constrained does the uncertainty on $h$ reduce to the 1-2\% level (still under the assumption of \lcdm). Of course, if the external data has systematics that shift the value of $\om$, then the value of $h$ derived from the GW plus this data will be biased. 

\begin{figure}
    \centering
    \includegraphics[width=\columnwidth]{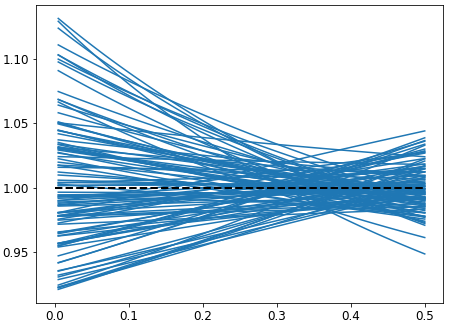}
    \caption{The distance uncertainty is shown as a scatter plot from a posterior predictive distribution sampling of the distances (relative to the input cosmology), for next generation GW data.} 
    \label{fig:10yr_ratio}
\end{figure}

Another, intuitive way of seeing the difficulty in GW sirens (or any 
cosmic distance) in determining cleanly $H_0$ is provided in 
Fig.~\ref{fig:10yr_ratio}.  We plot realizations of the ratio of GW luminosity distances for $\Lambda$CDM cosmologies with parameters drawn from the posterior relative to the input cosmology. 
The size of the scatter at different redshifts indicates at which redshifts the distance is better constrained. Both the parameter covariances and the number of events at a given redshift enter into the scatter. Unfortunately, redshift zero and hence $D_L(z\ll1)=H_0^{-1}z$ has large uncertainty. Thus we do not expect $H_0$ to be constrained near the 1\% level when we are appropriately agnostic about the expansion history.

\begin{figure}
    \centering
    \includegraphics[width=\columnwidth]{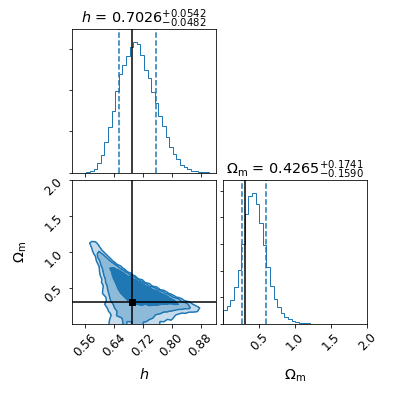}
    \includegraphics[width=\columnwidth]{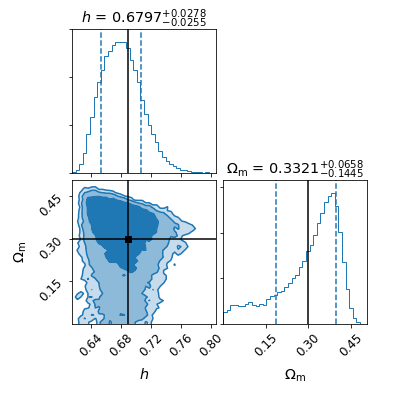}
    \caption{Forecast posterior for mock data generated from a $\Lambda$CDM cosmology, but marginalizing over $w_0$ and $w_a$ in the MCMC fit, for next generation (top) and next next generation (bottom) sensitivities.}
    \label{fig:w0wa_nearfuture}
\end{figure}

If we allow for standard $w_0$-$w_a$ dynamical dark energy freedom in the background we see an even bleaker picture. Now, even with the matter density perfectly known, GW sirens deliver only $\sigma(h)=0.039$, i.e.\ 6\%. With 
$\om$ as a fit parameter the covariance increases 
the MCMC uncertainty on $h$, even in the next next 
generation case. The 
MCMC contours in the $\om$--$h$ plane are shown in 
Fig.~\ref{fig:w0wa_nearfuture}. 
Thus the cosmological model assumed plays a critical role in the constraints from GW sirens at cosmic distances.

\section{Bias vs Cosmological Model} \label{sec:bias} 

Apart from the issue of precision on $H_0$, if insufficient freedom is given to 
the background cosmology fit then the resulting value of $H_0$ (and $\om$) will 
be biased. We explore the magnitude of this effect by choosing two alternative $w_0$--$w_a$ 
model points on the 68\% confidence contour of the current joint probe analysis 
in \cite{2018ApJ...859..101S} and generating GW data sets in these cosmologies. If 
these are then analyzed within the \lcdm\ model, parameter biases ensue. The two 
cosmologies used are ($w_0$,$w_a$)=($-0.90$,$-0.75$) and ($-1.14$,0.35), with 
$\om=0.3$, $h=0.69$, and by construction are consistent with current combined data 
sets. 

After generating our mock luminosity distance dataset from GW+EM observations, we then use MCMC sampling to infer the 1D parameter fits and 2D joint confidence contours for a $\Lambda$CDM model.  
This allows us to assess the bias induced by the incorrect background model assumption, 
and its significance relative to the statistical precision. 
The data quality will be particularly important for this last question: as the 
precision improves a given bias becomes more important. Therefore we study both 
next generation and next next generation GW data sets.

\begin{figure}
    \centering
    \includegraphics[width=\columnwidth]{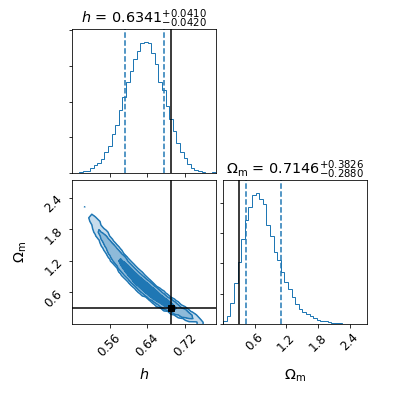}
    \includegraphics[width=\columnwidth]{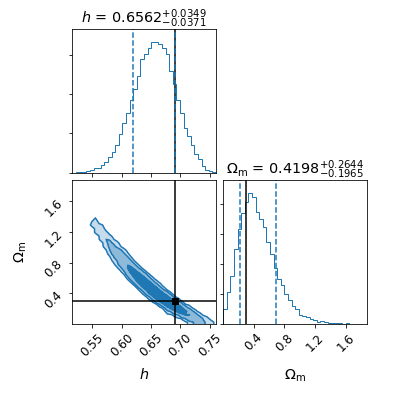}
    \caption{Forecast posteriors of the $\Lambda$CDM parameters for the next generation case, where the distances to the GW+EM events are generated from currently viable ($w_0$,$w_a$) cosmologies. The top panel is for $(w_0,w_a) = (-0.9,-0.75)$ and the bottom panel is for $(-1.14,0.35)$. 
    }
    \label{fig:10yr}
\end{figure}

We present the results for the next generation case in Fig.~\ref{fig:10yr}.  The 
precision obtained for the two input models is comparable, with $\sim$6\% uncertainty on 
$H_0$ and $\sim$50\% on $\om$. Clearly next generation GW alone will not give the 
desired constraint. The bias, due to misassuming \lcdm, shifts the fit contours so 
that the true input values are at the edge of 68\% confidence contour in each case. 

\begin{figure}
    \centering
    \includegraphics[width=\columnwidth]{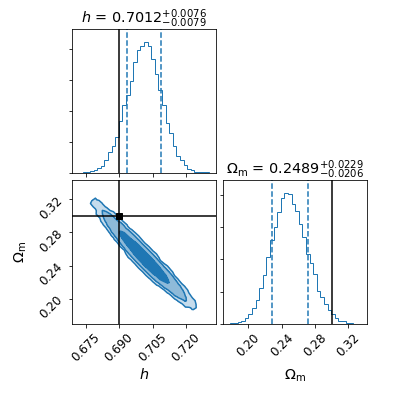}
    \includegraphics[width=\columnwidth]{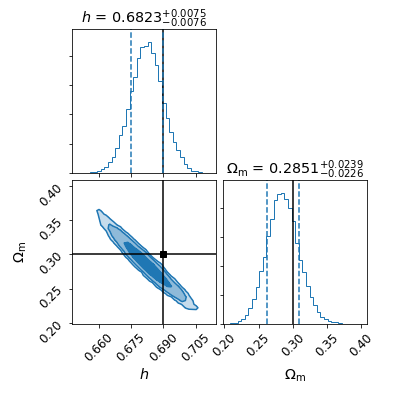}
    \caption{As Fig.~\ref{fig:10yr} but for next 
    next generation data.} 
    \label{fig:ff}
\end{figure}

Figure~\ref{fig:ff} repeats the analysis for the next next generation data case. 
The parameter precision is now strongly improved, to 1.1\% on $H_0$ and 8\% on 
$\om$. However, the bias is much more severe, with the true values lying outside 
the 99.7\% joint confidence contour, i.e.\ roughly $3\sigma$ bias (although the 1D 
values do not accurately show this tension). Note that knowing the actual value of matter density here from other observations could increase the bias in estimation of $H0$ due to our wrong assumption of the background expansion model.


\begin{figure}
    \centering
    \includegraphics[width=\columnwidth]{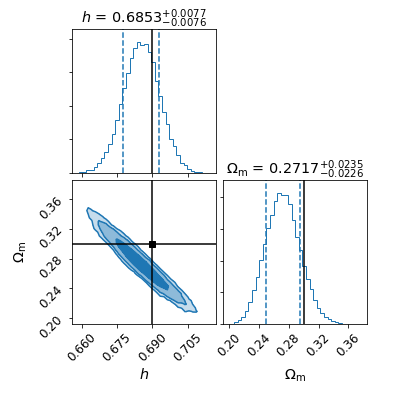}
    \caption{As Fig.~\ref{fig:ff} but for distances generated from a constant $w=-1.1$ cosmology.}
    \label{fig:const_w}
\end{figure}

The biases are evident even in a simple constant $w$ extension to \lcdm. If we 
take $w=-1.10$ from the 68\% confidence contour of the current joint data 
constraint of \cite{2018ApJ...859..101S}, Fig.~\ref{fig:const_w} shows the input values of  
($h$,$\,\om$) have been biased to outside the 99.7\% joint confidence contour in the \lcdm\ 
analysis (indeed to more than the equivalent of $5\sigma$).

From Eq.~\ref{eq:DL} and Fig.~\ref{fig:10yr_ratio} we can see that the bias must exist if the cosmological framework used in the fit does not cover the true 
cosmological model. For GW data to constrain primarily $H_0$, then $h(z)$ should be indistinguishable from 1 at the level of the statistical precision.  Similarly, if we go beyond $H_0$ to include $\Omega_{\rm m}$, then to fit these parameters without bias the  $h(z)$ for the \lcdm\ cosmology fit should be indistinguishable from the true  (potentially non-\lcdm) cosmology, again at the level of the statistical precision. 

The direction and magnitude of the parameter biases is a combination of the data 
properties (e.g.\ redshift distribution), cosmological parameter covariances, 
and parameter values. We have verified the MCMC results through the analytic 
Fisher bias formalism \citep{1998PhRvL..81.2004K,2006APh....26..102L}, which makes these dependencies more 
explicit. 
The bias on a parameter (such as $h$ or $\om$) is given by 
\bea 
\delta p_i&=&\left(F_{\rm sub}^{-1}\right)_{ij}\sum_k \frac{\partial{\mathcal O_k}}{\partial p_j}\frac{1}{\sigma_k^2}\Delta{\mathcal O_k}\\ 
&=&\left(F_{\rm sub}^{-1}\right)_{ij}\,\left[(1+w_0)F_{jw_0}+w_aF_{jw_a}\right]\ , 
\eea 
where $\Delta{\mathcal O_k}$ is the difference between the true distance and in the assumed model. 
Here the second line holds when the assumed cosmological model is a subset of a 
cosmological model with additional parameters, as \lcdm\ is a 
subset of a $w_0$--$w_a$ cosmology. 
This parameter bias estimation is in good agreement with the full Monte Carlo analysis we use.

\section{Discussions and Conclusions} \label{sec:concl} 

Gravitational wave sirens in conjunction with electromagnetic counterparts provide 
a new distance measure for the universe. New and improved detectors with better 
sensitivities and further redshift reach are exciting developments that will 
deliver abundant 
science from significant numbers of events. While GW provide an absolute
distance measurement they are not a panacea -- covariance between the evolution of 
the expansion $H(z)$ and the absolute scale today $H_0$ still exists. 

We have quantified how assuming that the expansion history $H(z)$ is purely of the 
$\Lambda$CDM form to infer the value of $H_0$ from GW luminosity distance data will yield inaccurate results should the true cosmology be different than $\Lambda$CDM. 
This holds even if the deviation from \lcdm\ cosmology is modest, within the  
68\% confidence level constraint from the current combination of data from several
probes. Indeed we find that even a constant $w$ model within the current 68\% joint confidence 
contour can deliver almost a $5\sigma$ bias if inappropriately analyzed within \lcdm. Fixing the value of the matter density $\om$ within \lcdm, as is sometimes 
adopted in predicting 1\% precision on $H_0$ from GW, is further problematic for accuracy. 

Next generation GW events will probe deeper into the universe, so all the freedom 
that enters into $H(z)$, from the imperfectly known matter density $\om$ and dark 
energy properties, will both dilute the precision and open 
up the potential for bias as we try to project distances to the very low redshift 
behavior involving only the Hubble constant $H_0$. Being properly agnostic about the
expansion history translates into uncertainties in $H_0$ that are well above 1\%. 
To quantify this, we carry out a Monte Carlo analysis simulating the GW+EM event 
distance data for next generation (roughly 2026) and next next generation experiments. 

Next generation data reaches 1.2\% precision on $H_0$ only if both \lcdm\ is 
assumed and $\om$ is perfectly known, with a degradation to 1.4\% if \lcdm\ is assumed 
and an external prior on $\om$ is used. For GW themselves, the precision is 5\% when 
restricted to \lcdm. Allowing for uncertainty in the cosmological model by including 
dynamical dark energy such as with $w_0$, $w_a$ dilutes the precision to 7\% -- barely more  constraining than the 
single local GW 
binary neutron star event already measured  \citep{2017Natur.551...85A,2017ApJ...848L..12A}. Thus cosmic GW data can clearly not be 
implemented as a pure prior on $H_0$. 

This is in no way a failing of GW data. Any cosmic distance measurement has the same issues 
with covariances (and note strong lensing time delays involve $H_0$ in a similar way 
to GW), and potential biases if unduly restricted to the wrong expansion 
model. 

The biases exist if fixing to the wrong $\Omega_{\rm m}$,  the wrong constant $w$, or in general assuming a wrong model or form of dark energy. (Note that the $w_0$--$w_a$ form does fit the distances out to $z=1$ to 0.1\% in a wide variety of viable models \citep{2008JCAP...10..042D}.) In this work we show that for the case of next next generation GW data we can have more than $3\sigma$ bias in estimation of $H_0$ while precision of the estimation can be very tight at 1.1\%. One safe approach to advocate is to carry out the analysis with model independent reconstruction techniques. 



If one could achieve substantial samples of GW+EM events at $z\lesssim0.05$ 
then most 
of the parameter covariance vanishes and one does get purer determination of $H_0$ 
independent of background cosmological model (modulo issues of peculiar velocities 
and coherent flows). This would be an exciting prospect, 
though event rates are currently too uncertain to obtain a clear estimate of the 
$H_0$ leverage.

\section*{Acknowledgements}

We thank KIAS, where this project was discussed among the co-authors, and especially Stephen Appleby for hospitality. A.S. would like to acknowledge the support of the National Research Foundation of Korea (NRF- 2016R1C1B2016478). EL is supported in part by the Energetic Cosmos Laboratory and by the U.S.\ Department of Energy, Office of Science, Office of High Energy Physics, under Award DE-SC-0007867 and contract no.\ DE-AC02- 05CH11231.




\bibliographystyle{mnras}
\bibliography{example2} 








\bsp	
\label{lastpage}
\end{document}